# Time-reversal symmetry breaking fractional quantum spin Hall insulator in moiré MoTe$_2$

Kaifei Kang[1*], Yichen Qiu[2], Bowen Shen[1], Kihong Lee[2], Zhengchao Xia[1], Yihang Zeng[2], Kenji Watanabe[3], Takashi Taniguchi[3], Jie Shan[1,2,4,5*], and Kin Fai Mak[1,2,4,5*]

[1] School of Applied and Engineering Physics, Cornell University, Ithaca, NY, USA
[2] Department of Physics, Cornell University, Ithaca, NY, USA
[3] National Institute for Materials Science, Tsukuba, Japan
[4] Kavli Institute at Cornell for Nanoscale Science, Ithaca, NY, USA
[5] Max Planck Institute for the Structure and Dynamics of Matter, Hamburg, Germany.

*Emails: kk726@cornell.edu; jie.shan@cornell.edu; kinfai.mak@cornell.edu

**Abstract**
Twisted bilayer transition metal dichalcogenide semiconductors, which support flat Chern bands with enhanced interaction effects, realize a platform for fractional Chern insulators and fractional quantum spin Hall (FQSH) insulators. A recent experiment has reported the emergence of a FQSH insulator protected by spin-$S_z$ conservation at a moiré lattice filling factor $v = 3$ in 2.1-degree twisted bilayer MoTe$_2$. Theoretical studies have proposed both time-reversal symmetric and asymmetric ground states as possible candidates for the observed FQSH insulator, but the nature of the state remains unexplored. Here we report the observation of spontaneous time-reversal symmetry breaking at generic fillings in 2.1-degree twisted bilayer MoTe$_2$ from $v < 1$ all the way to $v > 6$ except at $v = 2, 4$, and 6. Although zero Hall response is observed at $v = 3$ for magnetic fields higher than 20 mT, a finite anomalous Hall response accompanied by a magnetic hysteresis is observed at lower magnetic fields, demonstrating spontaneous time-reversal symmetry breaking. Our work shows the tendency towards ferromagnetism by doping the first three pairs of conjugate Chern bands in the material; it also sheds light on the nature of the FQSH insulator at $v = 3$.

**Main**

Moiré materials have realized a highly tunable platform for studying and engineering strongly correlated and topological phases of matter[1-5]. In AA-stacked twisted bilayer transition metal dichalcogenide (TMD) semiconductors, the electronic states near the valence band maximum originate from the K- and K'-valley states of monolayer TMDs[6-8], which are spin-split and spin-valley locked by a strong Ising spin-orbit field that defines a spin-$S_z$ quantization axis[9]. The emergence of a pseudospin skyrmion moiré lattice (Fig. 1a) endows the system with flat Chern bands of spin- and valley-contrasting Chern numbers in the moiré Brillouin zone[6-8]. Whereas early continuum model calculations have predicted Chern bands described by the Kane-Mele model at small twist angles[6-8], recent calculations taking into account lattice relaxations (in the moiré length scale) and multi-harmonic moiré couplings have refined the moiré band structure in the small twist angle limit[10-15], yielding multiple Chern bands of the same spin/valley Chern number ($C_s = +1$) in the K-valley (Fig. 1b) and $C_s = -1$ in the K'-valley. These flat Chern bands, which mimic a series of conjugate Landau levels related by time-reversal symmetry, provide a platform to study electron fractionalization in insulators under zero magnetic field[16-18].

Recent experiments on twisted bilayer TMDs have reported the observations of both the integer and fractional Chern insulators at hole-filling factors $\nu = 1$ and $\nu = \frac{2}{3}, \frac{3}{5}$, respectively[19-23]. Integer and fractional quantum spin Hall (QSH) insulators have also been reported at $\nu = 2, 4, 6$, and $\nu = 3$, respectively[24,25]. The emergence of QSH insulators is manifested by 1) a quantized conductance $\frac{\nu}{2}\frac{e^2}{h}$ per helical edge state pair from nonlocal transport measurements, 2) a zero Hall response, and 3) topological protection by spin-$S_z$ conservation. (Here $e$ and $h$ denote the electron charge and Planck's constant, respectively.) The last aspect is revealed by the observed insensitivity of helical edge state transport to an out-of-plane magnetic field (which preserves spin-$S_z$ conservation) and by a strong suppression of helical edge state transport even under a weak in-plane magnetic field (which breaks spin-$S_z$ conservation)[24,25]. The insensitivity to an out-of-plane magnetic field contrasts with a two-dimensional (2D) topological insulator protected by time-reversal symmetry[26], in which a magnetic field in any direction is expected to suppress helical edge state transport[27]. These results establish twisted bilayer TMDs as a platform for studying both the fractional Chern[28-42] and fractional quantum spin Hall (FQSH)[43-47] insulators.

Among the QSH insulators, the FQSH insulator at $\nu = 3$ carrying a fractionally quantized conductance $\frac{3}{2}\frac{e^2}{h}$ per helical edge state pair has attracted much recent theoretical interest[13,48-58]. Whereas a time-reversal symmetric ground state made of two decoupled spin-/valley-copies of $\nu = \frac{3}{2}$ fractional Chern insulators is a possible candidate, recent theoretical studies that consider the presence of intervalley Coulomb interactions have proposed other possible candidates with some breaking the time-reversal symmetry spontaneously[48,49,51,54,57,58]; examples include $Z_4$ topological order from intervalley pairing, vortex spin liquids, p+ip intervalley pairing states and Halperin particle-hole states, etc. Moreover, a recent exact diagonalization study on a pair of conjugate Landau levels has demonstrated the stability of a time-reversal symmetry breaking FQSH insulator at half Landau level filling[48]. (A time-reversal symmetry breaking QSH insulator is allowed due to spin-$S_z$ conservation[59-63].) Shedding light on the nature of the $\nu = 3$ insulating state is therefore a pressing experimental issue.

In this work, we report experimental evidence of spontaneous time-reversal symmetry breaking in the $\nu = 3$ FQSH insulator. Intriguingly, although zero Hall response is observed at a perpendicular magnetic field $B_\perp \gtrsim 20$ mT, a finite anomalous Hall response accompanied by a magnetic hysteresis is observed at $B_\perp \lesssim 20$ mT. We also observe a general tendency of the system towards ferromagnetism when the first three moiré valence bands are doped by holes. Our results constrain the possible ground states for the $\nu = 3$ FQSH insulator.

We perform magneto-transport and -optical measurements on a dual-gated device (Fig. 1a) of 2.1-degree twisted bilayer $MoTe_2$ (t$MoTe_2$) with four electrical contacts made to a rectangular region (about $2 \times 5$ μm$^2$), where a homogeneous moiré lattice is found. The material has a moiré period $a_M \approx 9.6$ nm and a moiré density $n_M \approx 1.26 \times 10^{12}$ cm$^{-2}$. The two gates allow independent tuning of the hole-filling factor $\nu$ and the perpendicular electric field $E$; the latter tunes the interlayer potential difference in t$MoTe_2$. To achieve good electrical contacts down to the mK temperature range, we use Platinum (Pt) as contact electrodes and monolayer $WSe_2$ as a tunneling barrier; we also define contact gates on top of the top gate to turn on the electrical contacts by heavily hole-doping the t$MoTe_2$ immediately adjacent to the Pt electrodes. We further define additional split gates to turn off any unwanted parallel conduction channels in the t$MoTe_2$ layer. Details on the device design have been reported in Ref. [24]. See Supplementary Materials for additional information on device fabrication and electrical measurements.

**Emergence of QSH insulators**
Figures 1c and 1d show the four-terminal longitudinal ($R_{xx}$) and Hall ($R_{xy}$) resistance, respectively, as a function of $\nu$ and $E$ at $B_\perp = 0.1$ T. Unless otherwise specified, the sample (lattice) temperature is at $T = 20$ mK. The inset of Fig. 1c shows the measurement geometry. The dashed lines mark the critical electric field $E_c$, above and below which holes are fully polarized to one of the $MoTe_2$ layers and shared between the layers, respectively. Selected line cuts from Fig. 1c and 1d near $E = 0$ are shown in Fig. 1e. For $E < E_c$, we observe 1) a vanishing $R_{xx}$ and a quantized $R_{xy} = \frac{h}{e^2}$ at $\nu = 1$; 2) a $R_{xy} = \frac{3}{2}\frac{h}{e^2}$ plateau at $\nu = \frac{2}{3}$ (Supplementary Fig. 1); 3) a $R_{xx}$ dip and a weak Hall plateau near $\nu = \frac{3}{2}$ (red arrows); and 4) dips in $R_{xx}$ and zero $R_{xy}$ at $\nu = 2, 3, 4,$ and 6. Figure 1e also shows the filling factor dependence of the two-terminal conductance ($G_{2t}$) at selected $E$'s measured with the geometry shown in the inset. Conductance plateaus at $G_{2t} \approx \nu \frac{e^2}{h}$ nearly independent of $E$ (for $E < E_c$) are observed at $\nu = 1, 2, 3, 4,$ and 6. In contrast, $G_{2t}$ for the compressible states away from these fillings is sensitive to changes in $E$.

As discussed in Ref. [24], the states at $\nu = 1$ and $\nu = \frac{2}{3}, \frac{3}{2}$, respectively, are integer and fractional Chern insulators that support chiral edge state transport; and the states at $\nu = 2, 3, 4,$ and 6 are QSH insulators that support helical edge state transport. Since two parallel edge channels are involved in the measurement geometry for $G_{2t}$ (Fig. 1e inset), the conductance per helical edge state pair is $\frac{\nu}{2}\frac{e^2}{h}$. The states at $\nu = 2, 4,$ and 6 are therefore single, double, and triple QSH insulators while the $\nu = 3$ state is a FQSH insulator with conductance $\frac{3}{2}\frac{e^2}{h}$ per helical edge state pair. The emergence of QSH insulators is also consistent with the nearly vanishing $R_{xx}$ at these fillings (Fig. 1c and 1e) because the measurement geometry in the inset of Fig. 1c is effectively a bridge circuit for the helical edge states. A small departure of $R_{xx}$ from zero at $\nu = 2, 3, 4,$ and 6 is a consequence of the imperfect conductance quantization for the helical edge states due to back scattering[64].

We compare the above observations with the continuum model band structure for 2.1-degree tMoTe$_2$ using the model parameters reported in Ref. [11] (see Supplementary Materials for details). Figure 1a shows the moiré potential. The maximum and minimum in the color map, which correspond to Te-Mo and Mo-Te stacking, respectively, constitute the two sublattice sites in a honeycomb moiré lattice[6]. The interlayer moiré potential (shown by a vector field) forms a pseudospin skyrmion moiré lattice[6] and generates a series of flat Chern bands (at least four) of the same Chern number $C_s = +1$ in the K-valley and another series of $C_s = -1$ in the K'-valley (Fig. 1b). The flat Chern bands, especially the second set, mimic a series of conjugate Landau levels related by time-reversal symmetry. The observed single, double, and triple QSH insulators at $\nu = 2, 4,$ and 6, respectively, are fully consistent with the calculated band structure. More than a pair of helical edge states is allowed due to spin-$S_z$ conservation[59-63].

**Emergence of ferromagnetism at generic fillings**

We now focus on the Hall response in the small electric field regime ($E < E_c$). (In general, $R_{xy}$ vanishes for $E > E_c$ as shown in Supplementary Fig. 2.) Figures 2a and 2b show $R_{xy}$ as a function of $\nu$ and $B_\perp$ at $E = 0$ and 30 mV/nm, respectively. Selected line cuts from Fig. 2a at fixed $\nu$'s are shown in Fig. 2c. (Due to the diverging sample resistance, $R_{xy}$ cannot be reliably measured for $\nu \lesssim 0.6$.) Small $R_{xy}$ is observed at $\nu = 2, 3, 4,$ and 6 for $B_\perp < 2$ T. The $\nu = 3$ state shows negligible dispersion with $B_\perp$, which is consistent with a FQSH insulator but inconsistent with a valley-polarized Chern insulator observed in larger twist angle samples[65,66]. Away from $\nu = 2, 3, 4,$ and 6, a sharp jump in $R_{xy}$ is observed at $B_\perp = 0$ T, suggesting the emergence of ferromagnetism and anomalous Hall effect at generic fillings, where the system is compressible. Moreover, whereas $R_{xy}$ is positive for $B_\perp > 0$ at most fillings, $R_{xy}$ turns negative in the window $2 < \nu < 3$.

Next, we investigate the magnetic field dependence of $R_{xx}$ and $R_{xy}$ in the low field regime at selected fillings in Fig. 3. Consistent with a Chern insulator near $\nu = 1$, we observe a quantized anomalous Hall effect and a clear magnetic hysteresis, accompanied by a vanishing $R_{xx}$ (except near the magnetic switching at $B_\perp \approx \pm 100$ mT). Anomalous Hall effect and magnetic hysteresis are also observed in $2 < \nu < 4$, i.e. hole doping the second pair of conjugate Chern bands. In contrast to $\nu < 2$, $R_{xy}$ is negative for $B_\perp > 0$ in $2 < \nu < 3$; it switches sign at $\nu = 3$ and remains positive in $3 < \nu < 4$. Intriguingly, a vanishing $R_{xy}$ is observed at $\nu = 3$ for $B_\perp \gtrsim 20$ mT (and persists to much higher field as shown in Fig. 2a and 2b) while a magnetic hysteresis and a finite anomalous Hall response is observed for $B_\perp \lesssim 20$ mT. Anomalous Hall effect and magnetic hysteresis are also observed for $4 < \nu < 6$, i.e. hole doping the third pair of conjugate Chern bands. The emergence of ferromagnetism at generic fillings away from $\nu = 2, 4,$ and 6 shows the presence of strong electronic correlations in at least the first three pairs of conjugate Chern bands in 2.1-degree tMoTe$_2$.

We also study the temperature dependence of the Hall response at selected fillings that cover the first three pairs of conjugate Chern bands in Fig. 4. Figure 4a shows the difference in $R_{xy}$ ($\Delta R_{xy}$) between forward and backward magnetic field scans as a function of $B_\perp$ and $T$ at different fillings; a nonzero $\Delta R_{xy}$ implies the presence of magnetic hysteresis or ferromagnetism. Both the coercive field and the zero-field $\Delta R_{xy}$ decrease continuously with increasing temperature. The latter vanishes at the critical temperature $T_c$ (Fig. 4b), which is about 2-3 K for hole doping the first two pairs of conjugate Chern bands (i.e. $\nu < 4$) and is about 1-2 K for doping the third pair of bands (i.e. $4 < \nu < 6$).

We further supplement the transport measurements with magnetic circular dichroism (MCD) spectroscopy. We performed MCD measurements on the same device but at a location slightly away from the transport channel, where it is not covered by the metallic contact and split gates (Fig. 5a); reliable MCD measurements cannot be achieved on regions covered by the metallic gates, which distort the polarization state of the measurement. The sample twist angle at the MCD location is about 0.2-0.3 degrees smaller than that of the transport channel. Figure 5b shows the spectrally integrated MCD (over the moiré exciton resonance of tMoTe$_2$, see Supplementary Materials) as a function of $\nu$ and $E$ at $B_\perp = 50$ mT and $T = 1.55$ K. As we will demonstrate, the spectrally integrated MCD (or MCD) under a small magnetic field is an indicator of spontaneous time-reversal symmetry breaking. MCD hot spots are observed near both $\nu = 1$ and 3 for $E < E_c$. We also obtained similar results on another device with a similar twist angle (2.2 degrees) in Supplementary Fig. 3.

Figure 5c examines the magnetic field-dependent MCD at varying fillings near $\nu = 3$ and $E = 0$. The curves are vertically displaced for clarity. Consistent with the transport measurements, magnetic hysteresis with a coercive field around 20 mT is observed at generic fillings, demonstrating spontaneous time-reversal symmetry breaking. In contrast to the Hall response, however, no sign change in MCD is observed at $\nu = 3$; instead, a similar response curve of varying amplitude is observed across $\nu = 3$. The inset shows the filling factor dependence of the size of the hysteresis (or spontaneous MCD), which is measured by the difference in MCD at $B_\perp = 0$ T between forward and backward field scans. Spontaneous MCD is observed for $\nu < 4$ with the strongest response near $\nu = 1$ and 3. The absence of spontaneous MCD for $\nu > 4$ is due to the elevated base temperature ($T = 1.55$ K) for the MCD measurements compared to the transport studies (see Fig. 4). We further study the temperature dependence of the MCD response at $\nu = 3$ in Fig. 5d. With increasing temperature, the magnetic hysteresis disappears gradually. The inset summarizes the temperature dependence of the spontaneous MCD, which vanishes at a $T_c \approx 4$ K.

**Discussions**
Overall, the MCD results corroborate the Hall data. However, the relative sign difference in MCD and $R_{xy}$ for doping across $\nu = 3$ demonstrates that the optical and the DC Hall conductivities need not provide the same information. Compared to the MCD results, which concern the optical transitions between the occupied and unoccupied states separated by a large optical gap, it is more straightforward to interpret the anomalous Hall data, which concern the low-energy response of the many-body ground states. In particular, because of the uniform Berry curvature distribution in the flat Chern bands[11], the anomalous Hall conductance, which is proportional to the enclosed Berry phase of the occupied states[67], is directly connected to the spin/valley occupancy of the flat bands. We will therefore focus our discussions on the anomalous Hall response and only use the MCD results as supplementary information.

The implications of the experimental results are summarized in Fig. 6 using the calculated band structure in Fig. 1b. First, the quantized $R_{xy} = \frac{h}{e^2}$ and $R_{xy} = \frac{3}{2}\frac{h}{e^2}$ at $\nu = 1$ and $\nu = \frac{2}{3}$, respectively (Fig. 1e and Supplementary Fig. 1), show that, for $\nu \leq 1$, holes are injected solely into the first Chern band from the K-valley ($C_s = +1$), as illustrated in Fig. 6a. The strong electron-electron interactions stabilized a spontaneously valley-polarized state here. In addition, the observation of a fractional Chern insulator with a weak Hall plateau and a $R_{xx}$ dip near $\nu = \frac{3}{2}$ (Fig. 1e) shows that hole injection immediately beyond $\nu = 1$ occupies the second K-valley band (rather than the first K'-valley band), as illustrated in Fig. 6b. Only doping beyond $\nu = \frac{3}{2}$ that holes start occupying the first K'-valley band and leaving the second K-valley band, leading to a quick suppression of $R_{xy}$ with doping (Fig. 1e). At $\nu = 2$, the first pair of conjugate Chern bands is fully occupied (Fig. 6c); a QSH insulator with $R_{xy} \approx 0$ emerges.

The effects of doping the second pair of conjugate bands (i.e. $2 \leq \nu \leq 4$) are different. Instead of occupying the second K-valley band by doping immediately beyond $\nu = 2$, the observed $R_{xy} < 0$ (for $B_\perp > 0$) in the window $2 < \nu < 3$ shows that holes occupy the second K'-valley band (Fig. 6d). (See Supplementary Fig. 5 for the Zeeman effect under high magnetic fields.) Only upon a further increase in doping that holes begin to occupy the second K-valley band. The pair of conjugate bands is half-filled at $\nu = 3$ (Fig. 6e); the state with $G_{2t} \approx 3\frac{e^2}{h}$ and $R_{xy} = 0$ is a FQSH insulator. In contrast to the QSH insulator at $\nu = 2$, the FQSH insulator is stabilized by strong electronic correlations[13,48-58]. On the other hand, doping in the window $3 < \nu < 4$ favors a higher occupation in the second K-valley band than its K'-counterpart (Fig. 6f), resulting in $R_{xy} > 0$. The sign-change in $R_{xy}$ by doping through $\nu = 3$ (Fig. 2 and 3) suggests a cancellation in the Hall response of opposite signs from the pair of half-filled conjugate Chern bands at $\nu = 3$. At $\nu = 4$, the second pair of conjugate bands is fully occupied (Fig. 6g) and a double QSH insulator emerges.

Similar to the first pair of conjugate bands, doping the third pair of conjugate bands populates the K-valley band more than the K'-valley band (Fig. 6h), resulting in $R_{xy} > 0$ in the window $4 < \nu < 6$. The two conjugate bands are fully occupied at $\nu = 6$ (Fig. 6i), corresponding to a triple QSH insulator. The absence of observable incompressible states and the lower $T_c$ in this region compared to the first two pairs of conjugate bands are consistent with an increased bandwidth (thus reduced electronic correlations) and a more significant screening effect here.

The contrasting behavior in the second pair of conjugate bands is likely connected to its very flat dispersion (Fig. 1b). As pointed out by Ref. [11], the second pair of conjugate bands has the most uniform Berry curvature distribution in the moiré Brillouin zone and is the closest in mimicking a perfectly flat Landau level, conditions of which favor the emergence of a FQSH insulator as shown by Ref. [48]. The observation of comparable $T_c$ for $2 < \nu < 4$ and for $\nu < 2$ (Fig. 4), albeit the screening effect is more significant in the case of $2 < \nu < 4$, supports the picture of a flatter second pair of conjugate bands with stronger electronic correlations compared to the first pair. However, exactly how the unique band dispersion and the strong electronic correlations combine to yield a spontaneously K'(rather than K)-polarized state in the window $2 < \nu < 3$ (Fig. 6d) requires further investigations.

For the FQSH insulator at $\nu = 3$, the hysteretic anomalous Hall response (for $B_\perp \lesssim 20$ mT, Fig. 3) and the spontaneous MCD (Fig. 5) demonstrate spontaneous time-reversal symmetry breaking. The observations are consistent with recent theoretical proposals of ground states that break time-reversal symmetry spontaneously at $\nu = 3$ (Ref. [48,49,51,54,57,58]). In particular, the Halperin particle-hole FQSH states that carry a finite valley polarization[54] are expected to support a zero Hall response but a finite MCD. Meanwhile, other time-reversal symmetry breaking FQSH states, such as the vortex spin liquids[57] and the Pfaffian/anti-Pfaffian pairs[48], could also support spontaneous MCD; additional theoretical studies are required to confirm this point.

Finally, we want to point out that the physics in the second pair of conjugate Chern bands is dependent on the twist angle. As shown in Supplementary Fig. 4, instead of a zero Hall response at $\nu = 3$, a $R_{xy}$ hot spot is observed near $\nu = 3$ in a 2.6-degree tMoTe$_2$ device, showing that the $\nu = 3$ state is no longer a QSH insulator, consistent with a recent report on a device with a similar twist angle[65]. As demonstrated by a recent exact diagonalization study on a pair of half-filled conjugate Landau levels[48], the FQSH insulator is a more stable ground state than a valley-polarized Chern insulator at $\nu = 3$ due to the weaker intervalley Coulomb repulsion compared to the intravalley counterpart. The flatter Chern bands in smaller twist angle samples, which can better mimic Landau levels, are thus expected to favor the FQSH insulator at $\nu = 3$ compared to larger twist angle samples. A more systematic twist angle dependence study around 2 degrees is required to provide a more complete picture.

**Conclusion**
To conclude, through magneto-transport and -optical studies, we observe a general tendency towards ferromagnetism by hole-doping the first three pairs of conjugate Chern bands in 2.1-degree tMoTe$_2$. In particular, we observe experimental evidence of spontaneous time-reversal symmetry breaking in the $\nu = 3$ FQSH insulator protected by spin-$S_z$ conservation. Our results support the proposed scenario of a time-reversal symmetry breaking FQSH insulator in a pair of half-filled conjugate flat Chern bands[48,49,51,54,57,58]; here spontaneous time-reversal symmetry breaking is favored because of the lower intervalley Coulomb energy cost compared to the time-reversal symmetric ground states[48,57]. Meanwhile, multiple candidates with broken time-reversal symmetry remain as possible ground states; additional experiments are required to further distinguish them.

**Acknowledgments**
We thank Ahmed Abouelkomsan, Gil Young Cho, Liang Fu, Chaoming Jian, Inti Sodemann, and Yahui Zhang for their helpful discussions.

**Figures**

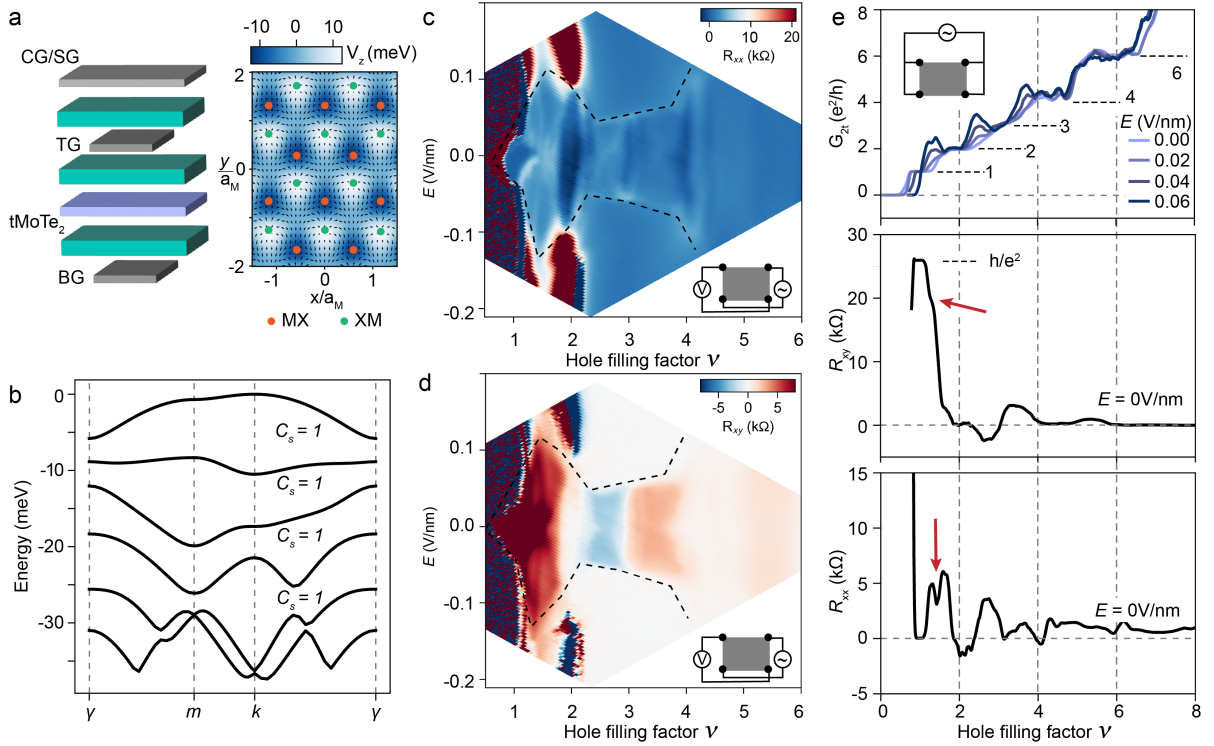

**Figure 1. QSH insulators in 2.1-degree tMoTe$_2$. a,** (Left) Schematic of a dual-gated device of 2.1-degree tMoTe$_2$. The top and bottom gates (TG and BG), which are made of hexagonal boron nitride and few-layer graphite, independently control the hole-filling factor $v$ and the perpendicular electric field $E$. The narrower top gate defines the transport channel. Contact and split gates (CG and SG) are further defined on top of TG to turn on the electrical contacts and turn off the parallel channels, respectively. (Right) Spatial dependence of the moiré potential $V_z \equiv [\Delta_b(\mathbf{r}) - \Delta_t(\mathbf{r})]/2$ and the 2D pseudospin field $\left[\text{Re}\left(\Delta_T^\dagger(\mathbf{r})\right), \text{Im}\left(\Delta_T^\dagger(\mathbf{r})\right)\right]$ (arrows) for 2.1-degree tMoTe$_2$ (see Supplementary Materials). The MX and XM sites are labeled (M = Mo and X = Te). **b,** Continuum model valence band structure in the moiré Brillouin zone in the K-valley. The first four flat Chern bands have the same spin/valley Chern number $C_s = +1$. **c, d,** The longitudinal ($R_{xx}$, **c**) and transverse ($R_{xy}$, **d**) resistance as a function of $v$ and $E$. The insets show the measurement geometry for the four-terminal device. The dashed lines mark the critical electric field $E_c$ that separates the regions with holes shared between the two MoTe$_2$ layers and holes fully polarized to one of the layers. **e,** Filling factor dependence of the two-terminal conductance ($G_{2t}$, top), $R_{xy}$ (middle) and $R_{xx}$ (bottom). The inset in the top panel shows the two-terminal measurement geometry. Conductance plateaus at $G_{2t} \approx v\frac{e^2}{h}$ (dashed lines) are observed at $v = 1, 2, 3, 4$ and $6$; vanishing $R_{xx}$ and $R_{xy}$ are observed at $v = 2, 3, 4$ and $6$. The arrows mark the fractional Chern insulator near $v = 3/2$. The results in **c-e** were acquired from the same device in Ref.[24] at a (lattice) temperature $T = 20$ mK and an out-of-plane magnetic field $B_\perp = 0.1$ T. Panel **e** is reproduced from Ref.[24].

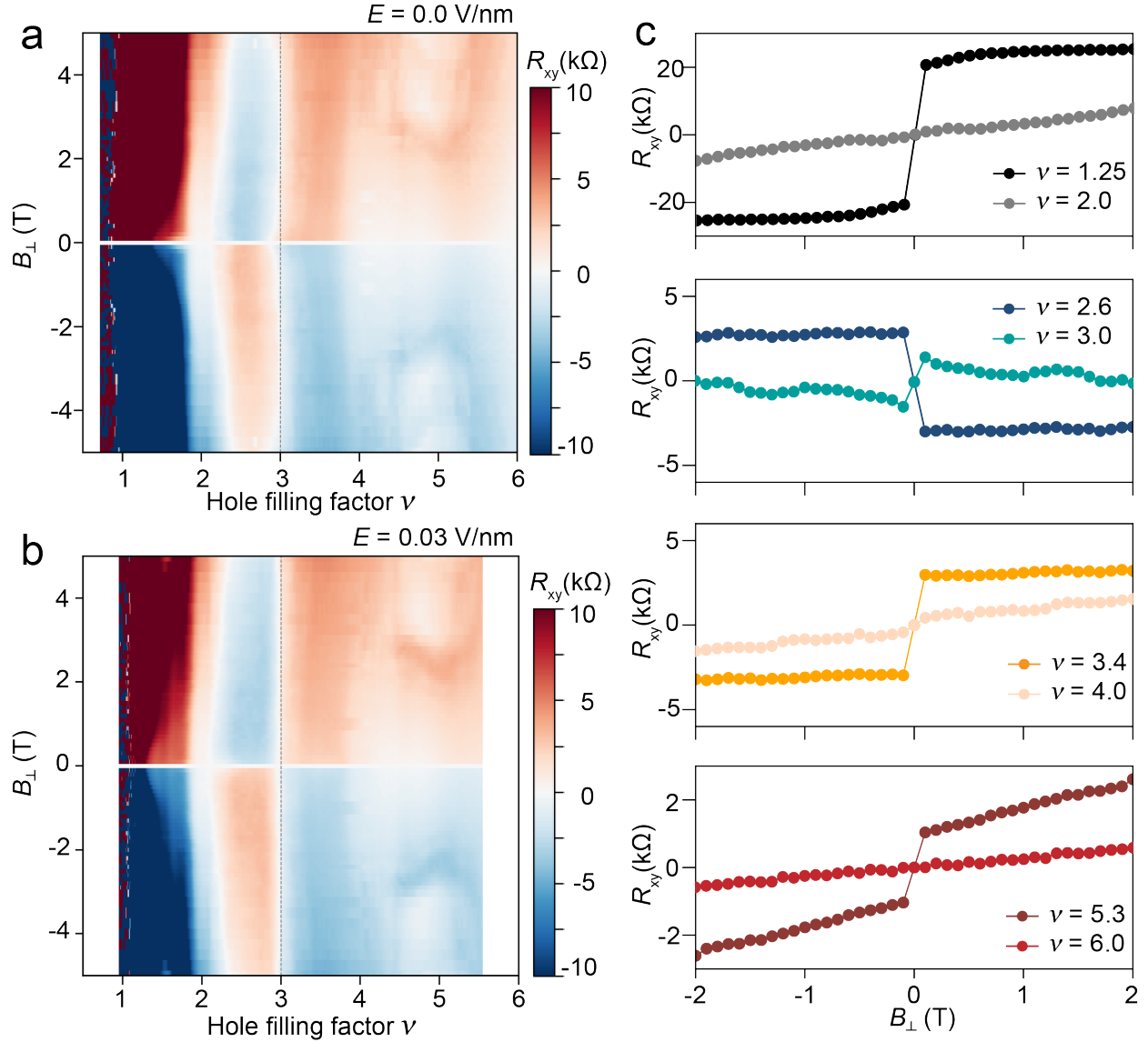

**Figure 2. Anomalous Hall response at generic fillings. a, b,** $R_{xy}$ as a function of $\nu$ and $B_\perp$ at $E = 0$ (**a**) and $E = 30$ mV/nm (**b**). The color scale is chosen to visualize the Hall response at $\nu \gtrsim 2$. The vertical dashed lines at $\nu = 3$ illustrate the absence of a dispersion of the state with $B_\perp$, consistent with a QSH insulator. **c,** Selected line cuts of **a** at varying filling factors. Except at $\nu = 2, 4$ and $6$, a $R_{xy}$ jump at $B_\perp = 0$ is observed at generic filling factors. The results were acquired at $T = 20$ mK.

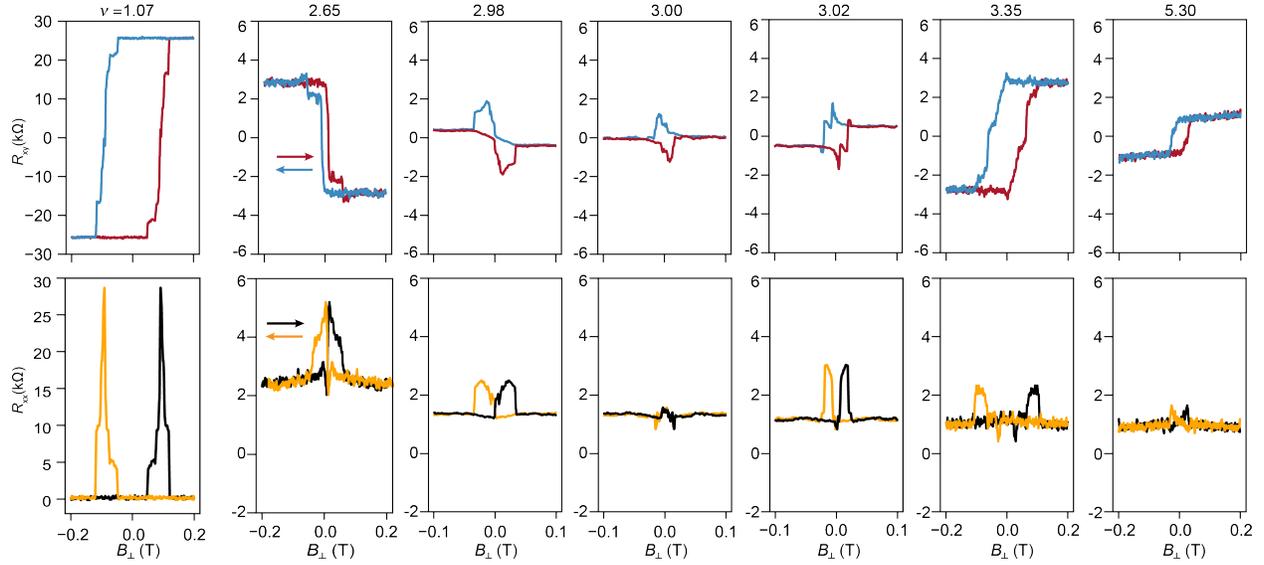

**Figure 3. Ferromagnetism at generic fillings and time-reversal symmetry breaking at $\nu = 3$.** Magnetic field dependence of $R_{xx}$ (bottom) and $R_{xy}$ (top) at selected filling factors that cover the first three pairs of conjugate Chern bands ($E \approx 0$ and $T = 20$ mK). Anomalous Hall effect and magnetic hysteresis are observed in all cases. A sign-change in $R_{xy}$ is observed at $\nu = 3$, where a hysteretic anomalous Hall response is observed at $B_\perp \lesssim 20$ mT but a zero Hall response is observed at $B_\perp \gtrsim 20$ mT.

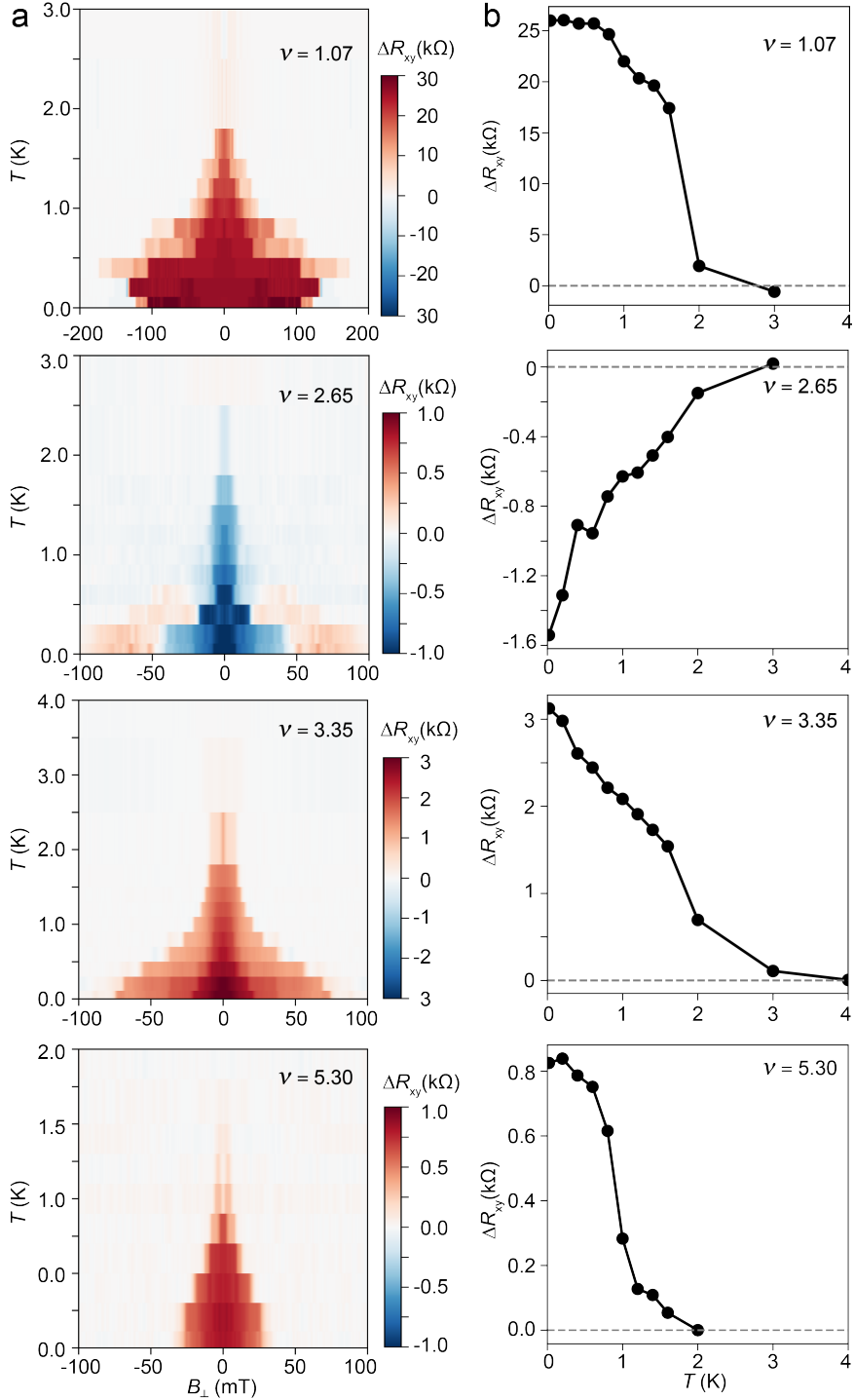

**Figure 4. Temperature dependence of the ferromagnetic ground states. a,** The difference in $R_{xy}$ ($\Delta R_{xy}$) between forward and backward magnetic field scan directions as a function of $B_\perp$ and $T$ at selected filling factors ($E \approx 0$). A nonzero $\Delta R_{xy}$ demonstrates the presence of ferromagnetism. The coercive field decreases monotonically with increasing temperature. **b,** Temperature dependence of $\Delta R_{xy}$ at $B_\perp = 0$ at the same filling factors. $\Delta R_{xy}$ vanishes at the magnetic critical temperature $T_c \approx$ 2-3 K.

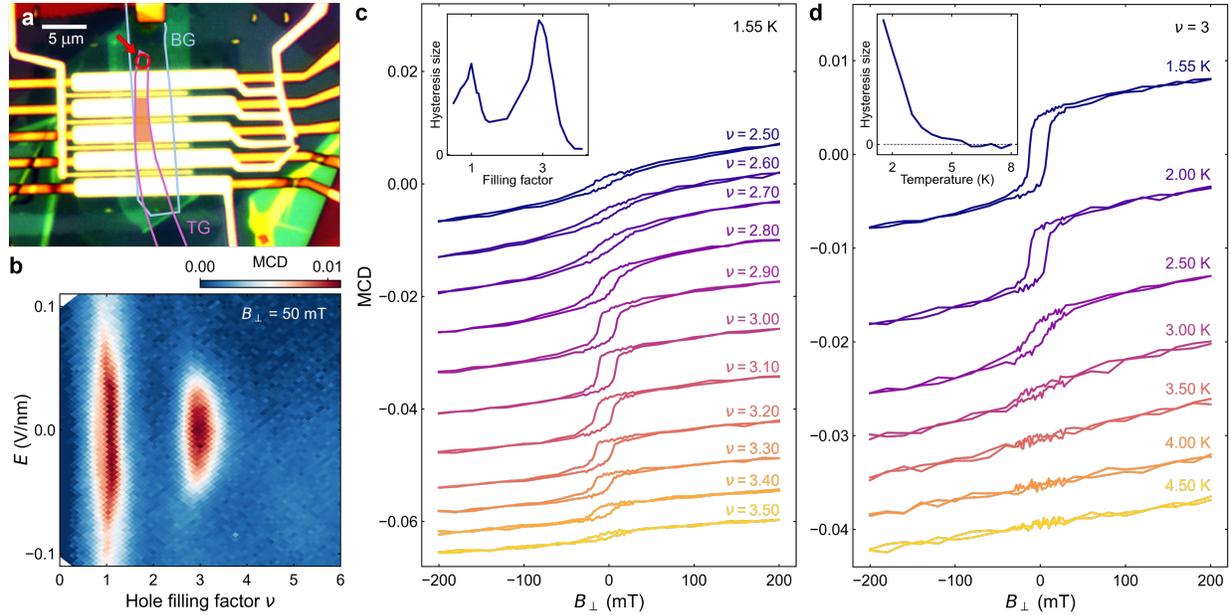

**Figure 5. MCD measurements. a,** Optical micrograph of the 2.1-degree tMoTe$_2$ device. The scale bar is 5 μm. The TG and BG graphite electrodes are labeled. The red-shaded area marks the 2.1-degree four-terminal device channel for the electrical transport measurements. The metallic electrodes are the contact and split gates (see Fig. 1a). The red arrow points to where the MCD measurements were performed; the location is chosen outside the metallic gates for reliable polarization-resolved measurements. **b,** Spectrally integrated MCD (or MCD) as a function of $\nu$ and $E$ at $B_\perp = 50$ mT and $T = 1.55$ K. Spontaneous MCD is observed near both $\nu = 1$ and 3 for $E < E_c$. **c,** Magnetic field dependent MCD at varying filling factors near $\nu = 3$ ($E \approx 0$ and $T = 1.55$ K). Data for both forward and backward field scan directions are shown. The curves at different fillings are vertically displaced for clarity. Ferromagnetism is observed. The inset shows the filling factor dependence of the spontaneous MCD at $T = 1.55$ K. **d,** Magnetic field dependent MCD at $\nu = 3$ ($E \approx 0$) with increasing temperature. The inset shows the temperature dependence of the spontaneous MCD; it vanishes at the critical temperature around 4 K.

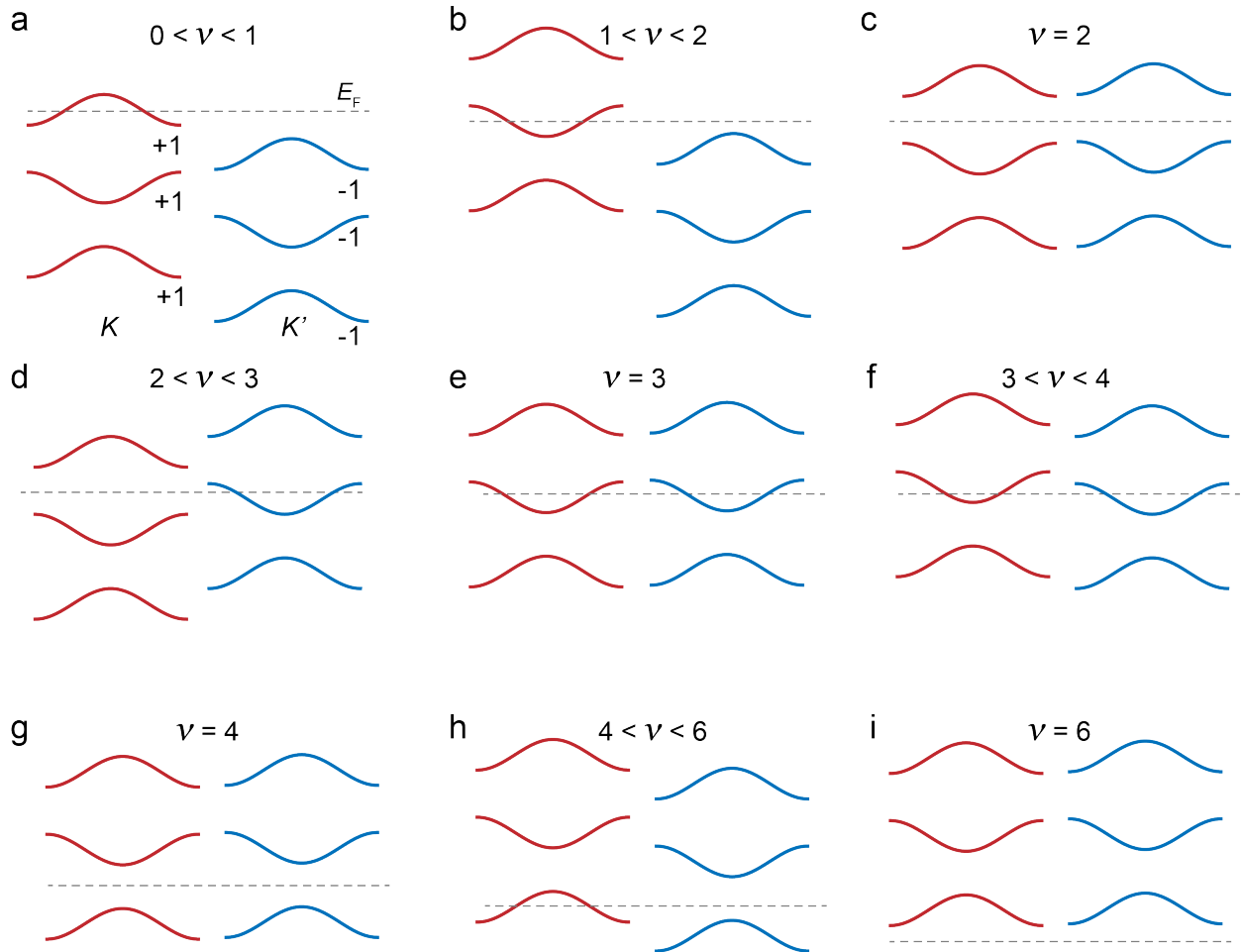

**Figure 6. Interaction physics in the first three pairs of conjugate Chern bands. a-i,** Schematic valley-resolved band alignments for $0 \leq \nu \leq 6$ deduced from the experimental anomalous Hall response. The spin/valley Chern numbers are labeled. The horizontal dashed line marks the Fermi level.

# Supplementary Materials for
# Time-reversal symmetry breaking fractional quantum spin Hall insulator in moiré MoTe$_2$

Kaifei Kang, Yichen Qiu, Bowen Shen, Kihong Lee, Zhengchao Xia, Yihang Zeng, Kenji Watanabe, Takashi Taniguchi, Jie Shan, and Kin Fai Mak

## Methods
**Device fabrication.**
Details of the device design and fabrication process have been reported in Ref. [24]. In short, we fabricated the dual-gated tMoTe$_2$ device using the tear-and-stack and layer-by-layer transfer method[68-70]. Flakes of MoTe$_2$, WSe$_2$, hexagonal boron nitride (hBN), and few-layer graphite were exfoliated from their bulk crystals onto SiO$_2$/Si substrates; their thicknesses were first estimated by their optical contrast, followed by direct measurements using atomic force microscopy. The exfoliated layers were then picked up by a polycarbonate thin film on PDMS (polydimethylsiloxane) in a sequence that produces the heterostructure illustrated in Fig. 1a. The finished stack was released onto a SiO$_2$/Si substrate with pre-patterned Pt electrodes at a temperature 180 degrees Celsius. Next, we defined the contact and split gates using electron-beam lithography. The top gate area was intentionally made smaller than the bottom gate area so that the device channel is defined by the top gate. The split gates turn off the tMoTe$_2$ regions only gated by the bottom gate so that they do not contribute to transport. The contact gates heavily hole-doped the tMoTe$_2$ regions adjacent to the Pt electrodes to achieve low contact resistance. We further annealed the device at 200 degrees Celsius under a high vacuum to improve the contacts. An optical micrograph of the finished device is shown in Fig. 5a.

To calibrate the twist angle of the device, we used the spin- and valley-polarized Landau levels emerging from $\nu = 2$ under high magnetic fields ($B_\perp > 7$ T), as reported in Ref. [24]. The density difference between two adjacent Landau levels is $\frac{B_\perp}{\Phi_0}$, where $\Phi_0 = \frac{h}{e}$ is the flux quantum. Using the known filling factor $\nu$, we can determine a twist angle (2.10±0.05) degrees and a moiré density $n_M \approx 1.26 \times 10^{12}$ cm$^{-2}$.

**Electrical measurements.**
We performed electrical transport measurements in a Bluefors LD250 dilution refrigerator equipped with a 12 T superconducting magnet using standard low-frequency (11.77 Hz) lock-in techniques. To measure the sample resistance, we biased the source-drain pair by a constant voltage (0.3 mV), which excites a bias current of less than 10 nA to avoid sample heating and/or high-bias effects. We used voltage preamplifiers with large input impedance (100 MΩ) to measure the four-terminal longitudinal and Hall resistances. To obtain $R_{xx}$ and $R_{xy}$, respectively, we symmetrized and antisymmetrized the resistances measured using the geometry in the inset of Fig. 1c under $B_\perp > 0$ and $B_\perp < 0$. The two-terminal resistance $R_{2t}$ was obtained by dividing the applied bias voltage (0.3 mV) by the measured bias current, followed by subtraction of a constant contact resistance independent of $\nu$ and $E$; the contact resistance was calibrated by $R_{xx}$ measurements at high doping densities ($\nu > 10$). (See Ref. [24] for additional details on contact resistance calibration). The two-terminal conductance is $G_{2t} = \frac{1}{R_{2t}}$.

**MCD measurements.**
We performed reflective magnetic circular dichroism (MCD) spectroscopy down to 1.6 K in a closed-cycle $^4$He cryostat (Attocube, Attodry 2100). Details of the measurement have been reported in Ref. [20]. In short, we used a superluminescent diode centered around 1070 nm as the excitation light source and a liquid nitrogen-cooled InGaAs array sensor as the detector. The incident light was collected by a microscope objective (numerical aperture 0.8) and focused onto the devices with a spot size of about 1.5 μm in diameter. We kept the incident light intensity below 20 nW/μm² to prevent heating or photo-doping. The MCD spectrum is defined as the difference in reflectance between the left- and right-handed circularly polarized light ($I^+$ and $I^-$, respectively), normalized by their sum: $\frac{I^+ - I^-}{I^+ + I^-}$. A strong enhancement of MCD near the attractive polaron resonance of MoTe$_2$ is observed. To obtain a spectrally integrated MCD signal, we integrated the absolute value of the MCD over a spectral window ranging from 1.121 to 1.133 eV that covers the attractive polaron resonance. The MCD results from two different devices with twist angles near 2 degrees are summarized in Fig. 5 and Supplementary Fig. 3.

**Continuum model band structure calculations.**
We calculated the moiré band structure of small-angle tMoTe$_2$ using a continuum model including the next-nearest neighbor interaction. We used the continuum model parameters from Ref. [11], which were obtained by fitting the continuum model bands to large-scale first-principles calculations. The Hamiltonian for the K valley states is given by

$$H_K = \begin{pmatrix} \frac{\hbar^2 k^2}{2m^*} + \Delta_t(\boldsymbol{r}) & \Delta_T(\boldsymbol{r}) \\ \Delta_T^\dagger(\boldsymbol{r}) & \frac{\hbar^2 k^2}{2m^*} + \Delta_b(\boldsymbol{r}) \end{pmatrix} \quad (1)$$

Here, $\frac{\hbar^2 k^2}{2m^*}$ is the kinetic energy with $k$ and $m^*$ denoting the moiré crystal momentum and effective mass; $\Delta_{t,b}(\boldsymbol{r}) = 2V_1 \sum_{j=1,3,5} \cos(\boldsymbol{g}_{n,j} \cdot \boldsymbol{r} \pm \psi_1) + 2V_2 \sum_{j=1,3,5} \cos(\boldsymbol{g}_{nn,j} \cdot \boldsymbol{r} \pm \psi_2)$ is the intralayer moiré potential for the top ($t$) and bottom ($b$) layer with first and second harmonic amplitudes of $V_1$ and $V_2$, and phases of $\psi_1$ and $\psi_2$ ($\boldsymbol{g}_{n,j}$ and $\boldsymbol{g}_{nn,j}$ are the moiré reciprocal wavevectors connecting nearest and next-nearest neighbor plane-wave basis in the same layer); $\Delta_T(\boldsymbol{r}) = w_1 \left(e^{i\boldsymbol{G}_{n,1}\cdot\boldsymbol{r}} + e^{i\boldsymbol{G}_{n,2}\cdot\boldsymbol{r}} + e^{i\boldsymbol{G}_{n,3}\cdot\boldsymbol{r}}\right) + w_2 \left(e^{i\boldsymbol{G}_{nn,1}\cdot\boldsymbol{r}} + e^{i\boldsymbol{G}_{nn,2}\cdot\boldsymbol{r}} + e^{i\boldsymbol{G}_{nn,3}\cdot\boldsymbol{r}}\right)$ is the interlayer interaction with first and second harmonic amplitudes of $w_1$ and $w_2$ ($\boldsymbol{G}_{n,j}$ and $\boldsymbol{G}_{nn,j}$ are the moiré reciprocal wavevector connecting nearest and next-nearest neighbor plane-wave basis in different layers). In the calculation, we used $m^* = 0.62\, m_0$ ($m_0$ denoting the free electron mass) and $(V_1, \psi_1, V_2, \psi_2, w_1, w_2) = (2.4\, meV, 90°, 1.0\, meV, 0°, -5.8\, meV, 2.8\, meV)$ (Ref. [11]). The Hamiltonian was diagonalized using the plane-wave method and cut off at the 5$^{th}$ shell of the moiré Brillouin zone. The spin/valley Chern number of each moiré band was obtained by integrating the Berry curvatures within the first Brillouin zone. The Hamiltonian for the K'-valley states is a time-reversed copy of Eqn. (1).

## Supplementary Figures

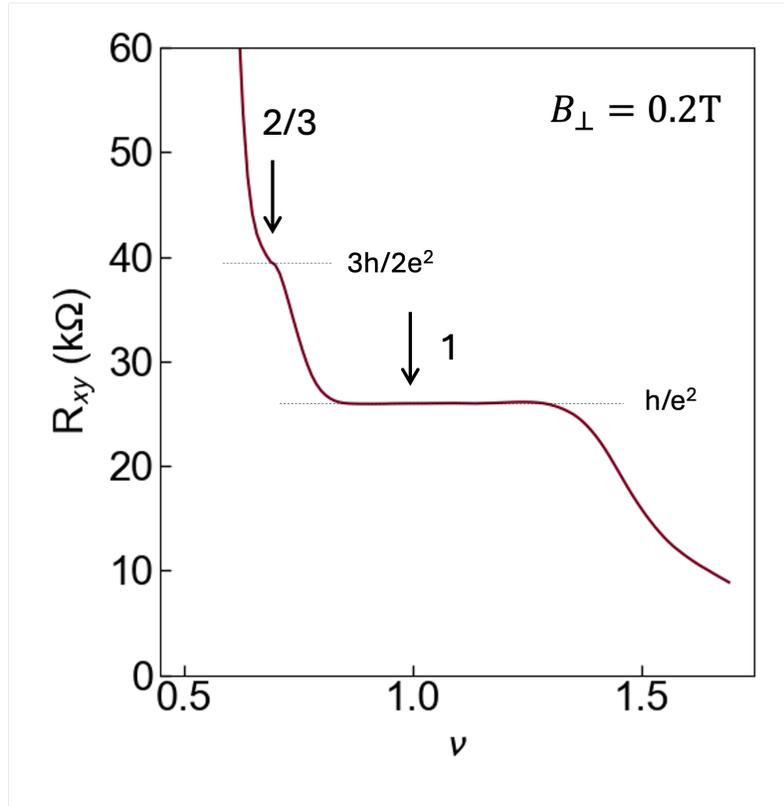

**Supplementary Figure 1. Fractional Chern insulator at $\nu = 2/3$.** Filling factor dependence of $R_{xy}$ near $E = 0$ ($B_\perp = 0.2$ T and $T = 20$ mK). In addition to the quantized anomalous Hall response $R_{xy} = \frac{h}{e^2}$ at $\nu = 1$, a fractionally quantized anomalous Hall response $R_{xy} = \frac{3h}{2e^2}$ is also observed at $\nu = 2/3$. Note that because of the higher magnetic field here than that applied in Fig. 1e, the fractional Chern insulator near $\nu = 3/2$ is overtaken by the integer Chern insulator at $\nu = 1$.

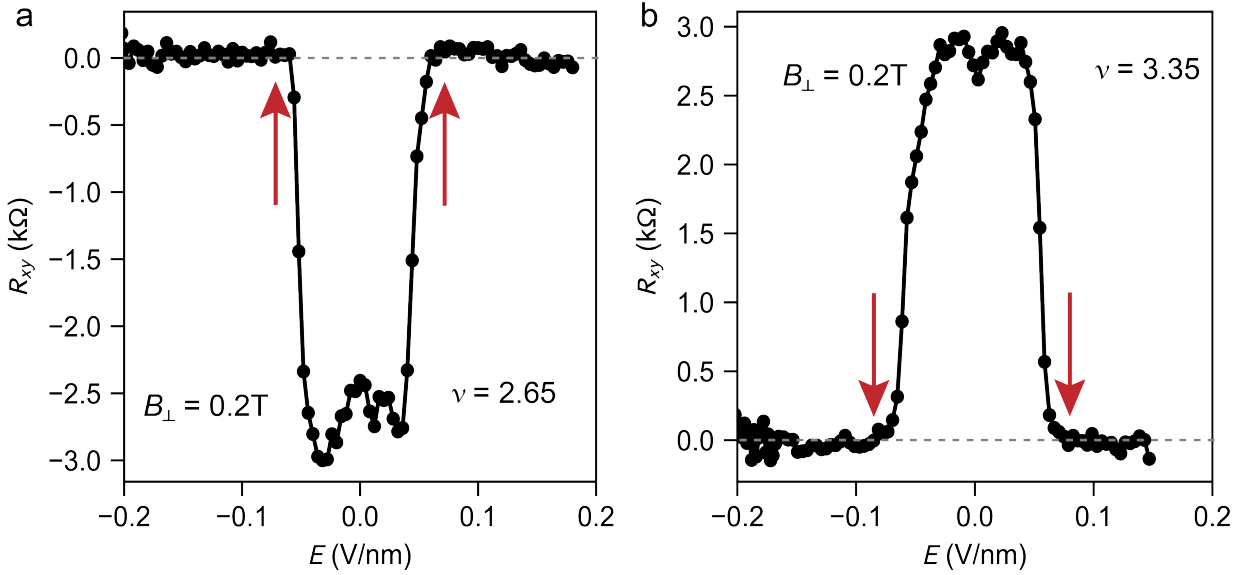

**Supplementary Figure 2. Electric field dependence of the Hall response. a, b,** Electric field dependence of $R_{xy}$ at $\nu = 2.65$ (**a**) and $\nu = 3.35$ (**b**) ($B_\perp = 0.2$ T and $T = 20$ mK). The anomalous Hall response vanishes at $E > E_c$, i.e. in the layer-polarized region of the phase diagram ($E_c$ marked by arrows).

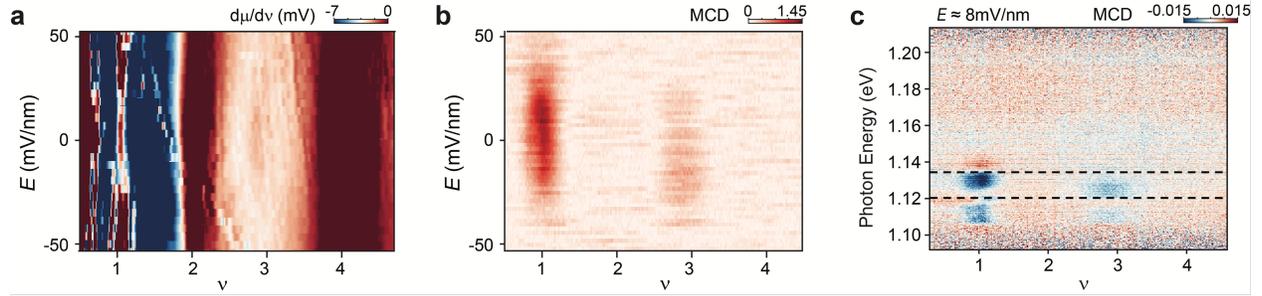

**Supplementary Figure 3. MCD results for a 2.2-degree device. a, b,** Electronic incompressibility (**a**) and spectrally integrated MCD (or MCD, **b**) as a function of $\nu$ and $E$ for a 2.2-degree tMoTe$_2$ device ($B_\perp = 50$ mT and $T = 1.6$ K). Incompressible states are observed at $\nu = 1, 2, 3$ and $4$. Spontaneous MCD or time-reversal symmetry breaking is observed near both $\nu = 1$ and $3$ for $E < E_c$. **c,** Filling factor dependence of the spontaneous MCD spectrum at $E \approx 8$ mV/nm. Resonance enhancement of MCD is observed near the charged exciton resonance of tMoTe$_2$. The dashed lines mark the spectral window for the MCD integration (see Methods).

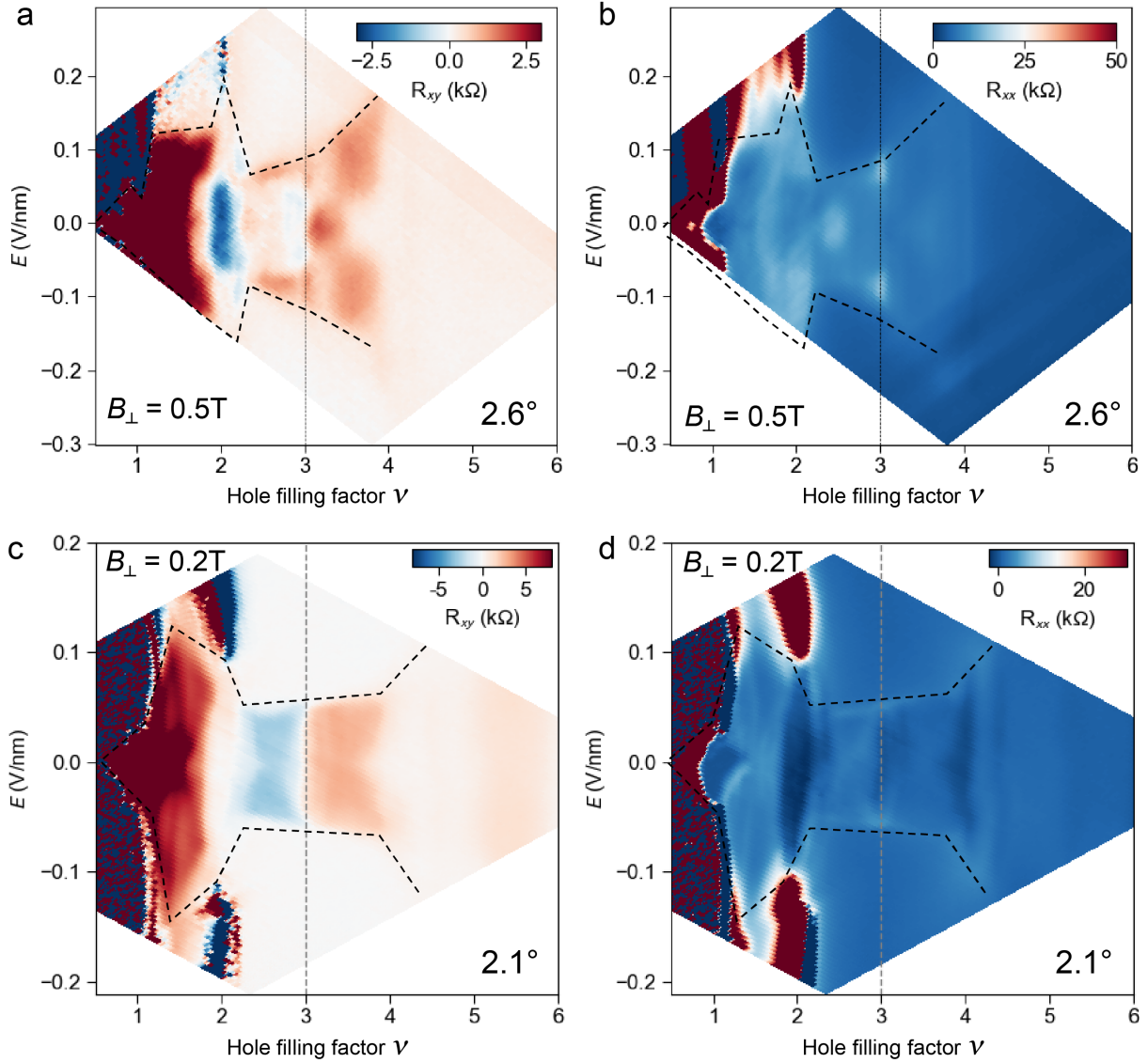

**Supplementary Figure 4. Illustration of twist angle dependence. a, b,** The transverse ($R_{xy}$, **a**) and longitudinal ($R_{xx}$, **b**) resistance as a function of $\nu$ and $E$ for a 2.6-degree tMoTe$_2$ device ($B_\perp = 0.5$ T and $T = 20$ mK). Similar data sets for the 2.1-degree device at $B_\perp = 0.2$ T and $T = 20$ mK are shown in **c** and **d** for comparison. The dashed lines mark the critical electric field $E_c$. The 2.6-degree data are largely consistent with those reported in Ref. [65]. In contrast to the 2.1-degree device, which shows zero Hall response at $\nu = 2$ and 3, a negative $R_{xy}$ and a $R_{xy}$ hot spot are observed at $\nu = 2$ and near $\nu = 3$, respectively, in the 2.6-degree device. In addition, $R_{xy}$ shows opposite signs in the window $2 < \nu < 3$ for the two devices. The results highlight the sensitivity of the physics in the second pair of conjugate Chern bands on the twist angle.

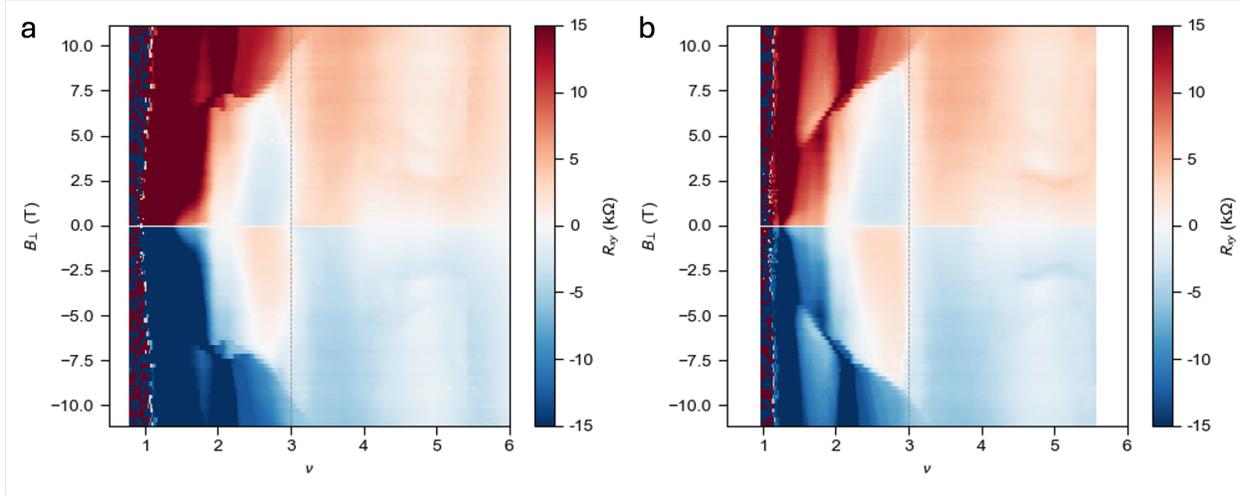

**Supplementary Figure 5. High-field Hall response. a, b,** $R_{xy}$ as a function of $\nu$ and $B_\perp$ up to high fields at $E = 0$ (**a**) and $E = 30$ mV/nm (**b**). The color scale is chosen to visualize the Hall response at $\nu \gtrsim 2$. The sign change in $R_{xy}$ in the window $2 < \nu < 3$ disappears at $B_\perp > 7 - 8$ T. The result shows that while the second K'-valley band is populated by holes in the window $2 < \nu < 3$ at low fields (Fig. 6d), the holes are polarized to the K-valley band at sufficiently high fields ($B_\perp > 7 - 8$ T) due to the Zeeman effect.